\documentclass[runningheads,openany]{svmult}
\usepackage[T1]{fontenc}
\usepackage{palatino}
\usepackage{euler}
\usepackage{graphicx}  
\usepackage{physprbb}  
\usepackage{bg2016}  

\def\half{{\textstyle{1\over2}}}
\def\thalf{{\textstyle{3\over2}}}

\def\i{{\rm i}}
\def\d{{\rm d}}

\def\h{{\scriptscriptstyle{1\over2}}}
\def\th{{\scriptscriptstyle{3\over2}}}
\def\CG#1#2#3#4#5#6{C^{#5#6}_{#1#2#3#4}}

\begin{document}
\title*{Puzzles in eta photoproduction: the 1685~MeV narrow peak\thanks{%
Talk delivered by B. Golli}
}

\author{B. Golli$^{a,c}$ and S. \v{S}irca$^{b,c}$
}
\authorrunning{B. Golli and S. \v{S}irca
}
\institute{%
$^a$
{Faculty of Education,
              University of Ljubljana,
              1000 Ljubljana, Slovenia}
\\
$^b${Faculty of Mathematics and Physics,
              University of Ljubljana,
              1000 Ljubljana, Slovenia}\\
$^c$
 and
{Jo\v{z}ef Stefan Institute, 
              1000 Ljubljana, Slovenia}
}

\titlerunning{Puzzles in eta photoproduction  \ldots}

\maketitle

\begin{abstract}
We claim that a narrow peak in the cross section near 1685~MeV 
in the $\gamma n\to \eta n$ channel can be explained through a 
peculiar radial behaviour of the $p$-wave quark states with $j=1/2$ 
and $j=3/2$ in the low lying S11 resonances and the opening of 
the $K\Sigma$ threshold rather than by an exotic resonance.
We explain the mechanism of its formation in the
framework of a coupled channel formalism which incorporates 
quasi-bound quark-model states corresponding to the two 
low lying resonances in the S11 partial wave.
A relation to the Single Quark Transition Model is pointed out.
\end{abstract}

\section{Motivation}

In this contribution we discuss a possible quark-model explanation 
for a narrow structure at $W\approx1685$~MeV in the 
$\gamma n\to\eta n$ reaction observed by the 
GRAAL Collaboration~\cite{Kuznetsov07} which, however, 
turned out to be absent in the $\eta p$ channel.
Azimov {\em et al.}~\cite{Azimov05} were the first to discuss the 
possibility that the structure could belong to a partner 
of the $\Theta^+$ pentaquark in the exotic antidecouplet of baryons.
More conventional explanations have attributed the peak to the 
threshold effect of the $K\Sigma$ channel~\cite{Doering10},
interference of the nearby $S_{11}$, $P_{11}$ and $P_{13}$
resonances~\cite{Shyam08}, constructive and destructive 
interference of the two lowest $S_{11}$ resonances in the $\eta n$ 
and $\eta p$ channels, respectively, as anticipated in the 
framework of the Giessen model~\cite{Shklyar07,Shklyar13} 
as well as in the Bonn-Gatchina analysis~\cite{BoGa09,BoGa15}.
In the framework of the constituent-quark model coupled to the 
pseudoscalar meson octet the (non)appearance of the peak was 
related to different EM multipoles (at the quark level) responsible 
for excitation in either of the two channels~\cite{Zhong11}.

\section{The coupled channel approach}

In our recent paper \cite{EPJ2016} we have systematically analysed 
the partial waves with sizable contributions to the $\eta N$, 
$K\Lambda$  and $K\Sigma$ decay channels using a SU(3) extended 
version of the Cloudy Bag Model (CBM)~\cite{Thomas85} 
which includes also the  $\rho$ and $\omega$ mesons\footnote{%
The method has been described in detail in our previous
papers~\cite{EPJ2005,EPJ2008,EPJ2009,EPJ2011,EPJ2013}
where we have analysed the scattering and 
electro-production amplitudes in different partial waves.}.
We have found that the main contribution to $\eta$
photoproduction at low and intermediate energies comes
from the S11 partial wave.
In this contribution we therefore concentrate on the S11 partial wave
in which the considered phenomenon is most clearly visible.

In our approach the main contribution to $\eta$ production in 
the S11 partial wave stems from the resonant part of the 
electroproduction amplitude which can be cast in the form
\begin{equation}
{\mathcal{M}_{\eta N\,\gamma N}^\mathrm{res}}  =
\sqrt{\omega_\gamma E_N^\gamma \over \omega_\pi E_N }\,
{\xi\over\pi{\cal V}_{N\mathcal{R}}^\pi}\,
  \langle{\Psi}_{\mathcal{R}}|{V}_\gamma
                |\Psi_N\rangle\, {T_{\eta N\,\pi N}} \>,
\label{Vgamma2M}
\end{equation}
where $T_{\eta N\,\pi N}$ is the $T$-matrix element pertinent
to the $\pi N\to \eta N$ channel,
$V_\gamma$ describes the interaction of the photon with the 
electromagnetic current and $\xi$ is the spin-isospin factor
depending on the considered multipole and the spin and isospin
of the outgoing hadrons.
Here
$|\Psi_{\mathcal{R}}\rangle=c_1(W)|N(1535)\rangle+c_2(W)|N(1650)\rangle$
with
\begin{eqnarray*}
|N(1535)\rangle &=& 
   \cos\vartheta|{\bf 70}, {}^2{\bf 8}, J=\half\rangle
-  \sin\vartheta|{\bf 70}, {}^4{\bf 8}, J=\half\rangle \>,
\\
|N(1650)\rangle &=& 
   \sin\vartheta|{\bf 70}, {}^2{\bf 8}, J=\half\rangle
+  \cos\vartheta|{\bf 70}, {}^4{\bf 8}, J=\half\rangle \>
\end{eqnarray*}
and $c_i(W)$ are $W$-dependent coefficients determined
in the coupled-channel calculation for scattering.

\begin{figure}[h!]
\hbox to\hsize{\kern-6pt
\includegraphics[width=63mm]{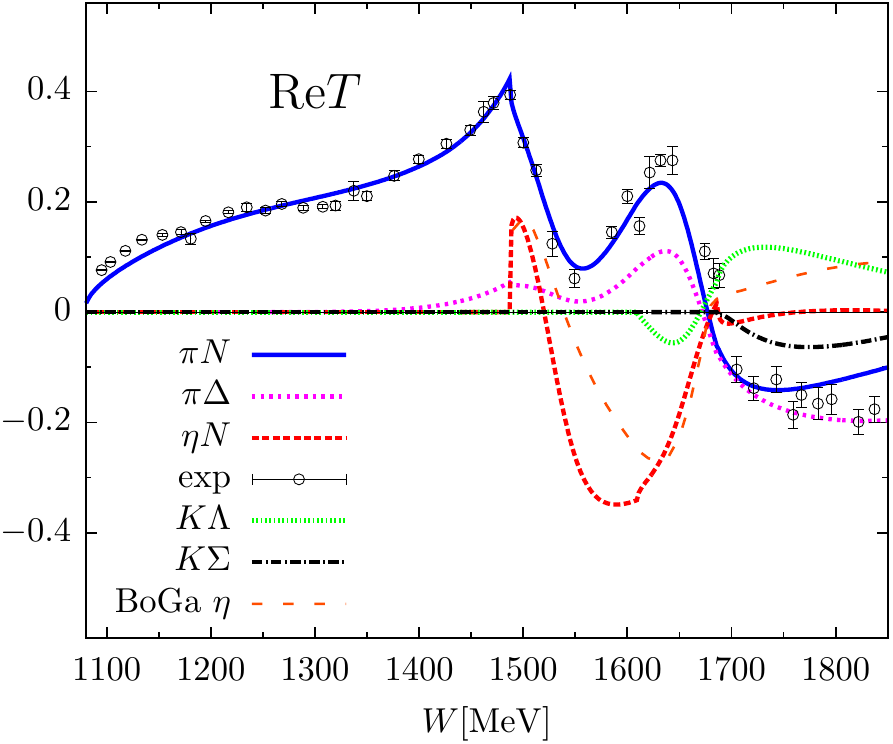}
\includegraphics[width=63mm]{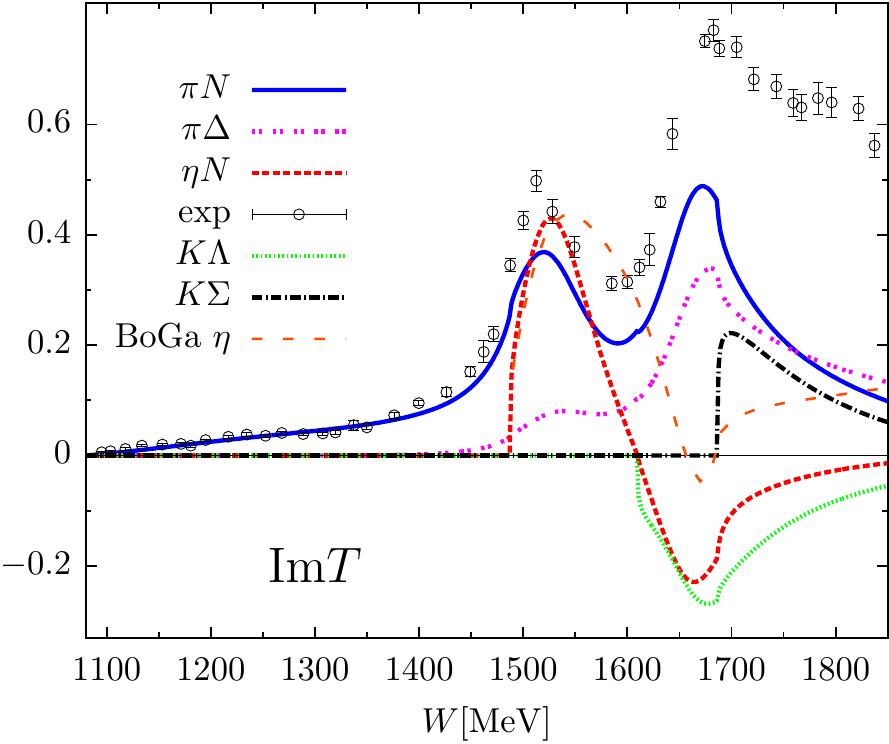}}
\caption{The real and imaginary parts of the scattering $T$ matrix 
for the dominant $\pi N$, $\pi\Delta$, $\eta N$, $K \Lambda$  
and $K\Sigma$ channels  in the $S_{11}$ partial wave.
The corresponding thin curve denote the 2014-2 solution of the 
Bonn-Gatchina group~\cite{BoGa-data} for the $\eta N$ channel.. 
The data points for the elastic channel are from the SAID 
partial-wave analysis \cite{bg:SAID}.}
\label{fig:TS11}
\end{figure}

The strong $T_{\eta N\,\pi N}$ amplitude is obtained in a coupled 
channel calculation with ten channels involving $\pi$, $\rho$, 
$\omega$, $\eta$ and $K$ mesons.
The most important channels are shown in Fig.~\ref{fig:TS11}.
The behaviour of the amplitudes is dominated by the 
$N(1535)$ and $N(1650)$ resonances as well as the $\eta N$, 
$K\Lambda$ and $K\Sigma$ thresholds.
In the present calculation we put the mixing angle $\theta$ to the 
popular value of $30^\circ$ and assume that all meson-quark coupling 
constants are fixed at their quark-model values dictated by the 
SU(3) symmetry.
While the real part of the elastic amplitude is well reproduced,
the imaginary part is rather strongly underestimated in the
region of the second resonance which can be to some extent 
attributed to  too strong couplings in the $\pi\Delta$, $K\Lambda$
and $K\Sigma$ channels.
This discrepancy should be taken into account when assessing
the quality of the photoproduction amplitudes in the following.

\section{The $\eta$ photoproduction amplitudes}

The electromagnetic amplitude in (\ref{Vgamma2M}) in the S11 
partial wave is dominated by the photon-quark coupling while 
the coupling to the pion cloud turns out to be small.
The spin doublet and quadruplet states involve quarks excited to 
the $p$ orbit with either $j=\half$ or $j=\thalf$~\cite{Myhrer84b}:
\begin{eqnarray}
|{}^4{\bf 8}_\h\rangle  &=& 
                \phantom{-}{1\over3} \,|(1s)^2(1p_{3/2})^1\rangle 
               + {\sqrt8\over 3} \,|(1s)^2(1p_{1/2})^1 \rangle \>,
\label{bg:84}\\
|{}^2{\bf 8}_\h\rangle  &=& 
               -{2\over3} \,|(1s)^2(1p_{3/2})^1\rangle 
               + {\sqrt2\over 6}\,|(1s)^2(1p_{1/2})^1\rangle 
               + {\sqrt2\over 2}\,|(1s)^2(1p_{1/2})^1\rangle' \>,
\label{bg:82}
\end{eqnarray}
where the last two components with $p_{1/2}$ correspond to 
coupling the two $s$-quarks to spin 1 and 0, respectively;
the flavour (isospin) part is not written explicitly. 
The quark part of the dominant $E_{0+}$ transition operator
can be cast in the form
\begin{eqnarray}
\int\d\vec{r}\,\vec{j}^q\cdot\vec{A}_{11}^e
&=& 
  \i\sum_{i=1}^3\left[\mathcal{M}_\h\Sigma_{11}^{[\h\h]}(i) + 
                     \mathcal{M}_\th\Sigma_{11}^{[\th\h]}(i)\right]
      \left[{\textstyle{1\over6} + {1\over2}}\,\tau_0(i)\right],
\label{bg:VE1}
\end{eqnarray}
where
\begin{eqnarray}
\mathcal{M}_\h&=&  \sqrt{2\over3}\,
\int\d r\,r^2 \left[j_0(qr)\left(3v^p_\h(r)u^s(r) +u^p_\h(r) v^s(r)\right)
                         -2j_2(qr)u^p_\h(r) v^s(r)\right],\>\>
\label{bg:M12}\\
\mathcal{M}_\th&=& \sqrt{2\over3}\,
\int\d r\,r^2 \left[2j_0(qr)u^p_\th(r) v^s(r)
      +\half j_2(qr)\left(u^p_\th(r) v^s(r)-3v^p_\th(r) u^s(r)\right)
     \right]\!.\kern15pt
\label{bg:M32}
\end{eqnarray}
The quark transition operator is defined through
$\langle ljm_j|\Sigma^{[j\h]}_{LM}|\half m_s\rangle = 
\CG{\h}{m_s}{L}{M}{j}{m_j}$.

Evaluating (\ref{bg:VE1}) between the resonant states and the 
nucleon we notice that for the proton, the isoscalar part of the 
charge operator exactly cancels the isovector part in the case 
of the first two components in (\ref{bg:84}) and (\ref{bg:82}).
This is a general property known as the Moorhouse selection 
rule~\cite{Moorhouse1966} and follows from the fact that the
flavour part in these two components corresponds to the
mixed symmetric state $\phi_{\mathrm{M,S}}$.
The proton therefore receives no contribution 
from the $1s\to 1p_{3/2}$  transition. 
This is not the case with the neutron which receives contributions
from all components in (\ref{bg:84}) and (\ref{bg:82}).
The quark in the $1p_{3/2}$ orbit has a distinctly different radial
behaviour  from that in the  $1p_{1/2}$ orbit, which is reflected 
in a different $q$- and $W$-behaviour of the amplitudes 
(\ref{bg:M12}) and (\ref{bg:M32}).

\begin{figure}[h!]
\hbox to\hsize{\kern-1mm
\includegraphics[width=62.3mm]{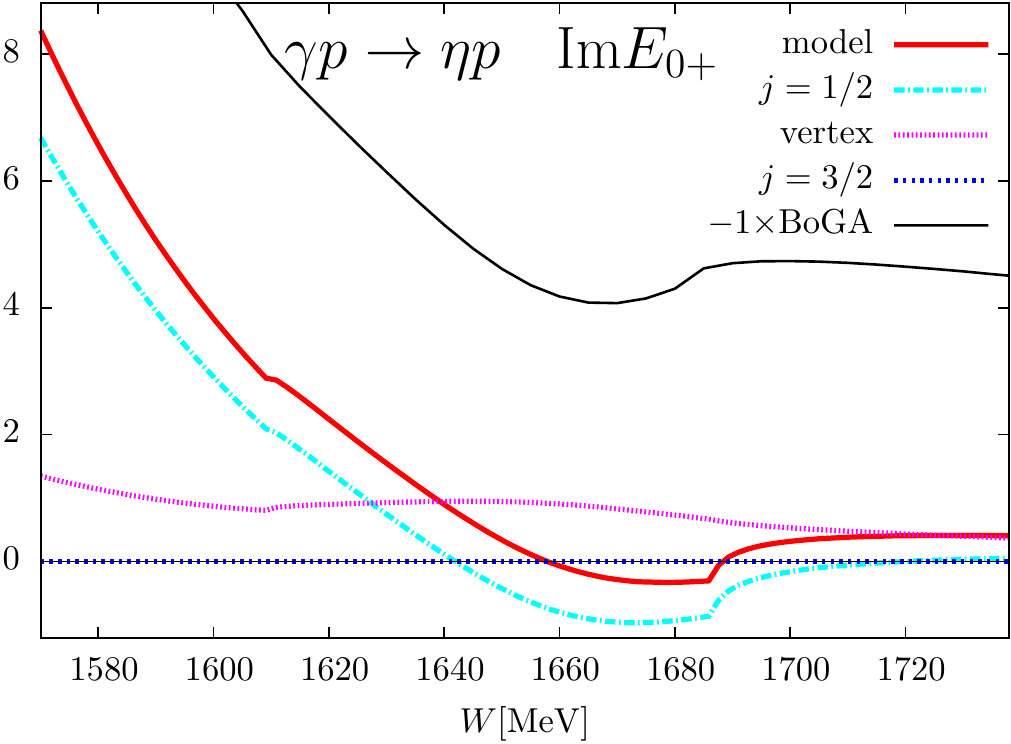}\kern2mm
\includegraphics[width=63mm]{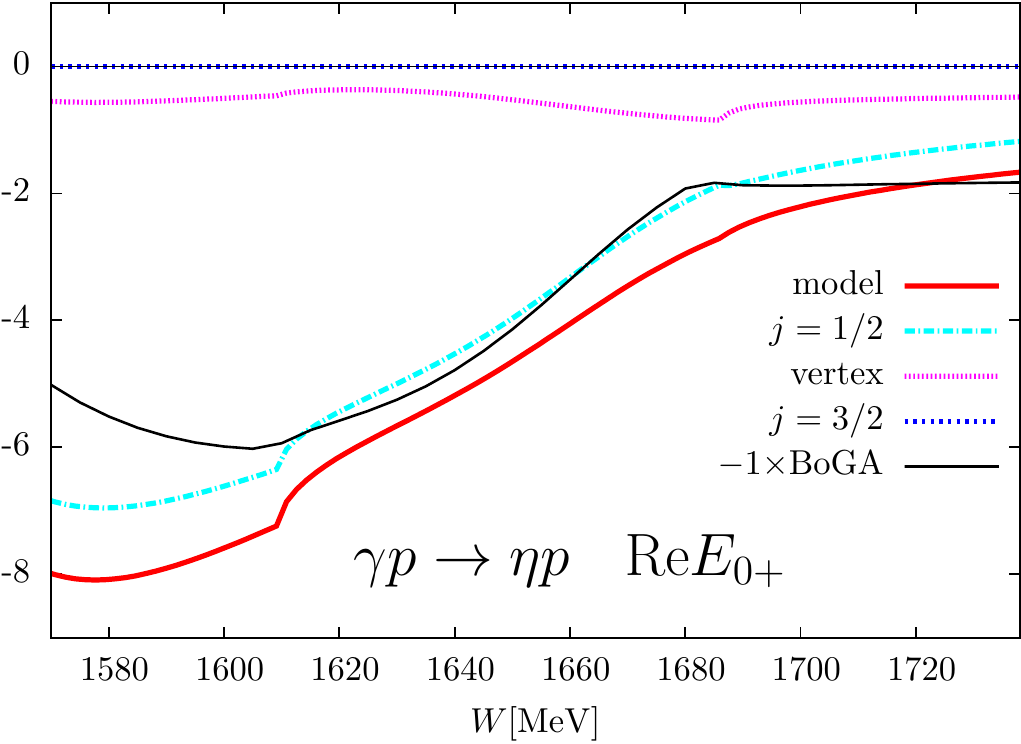}}
\hbox to\hsize{\kern-1.5mm
\includegraphics[width=63mm]{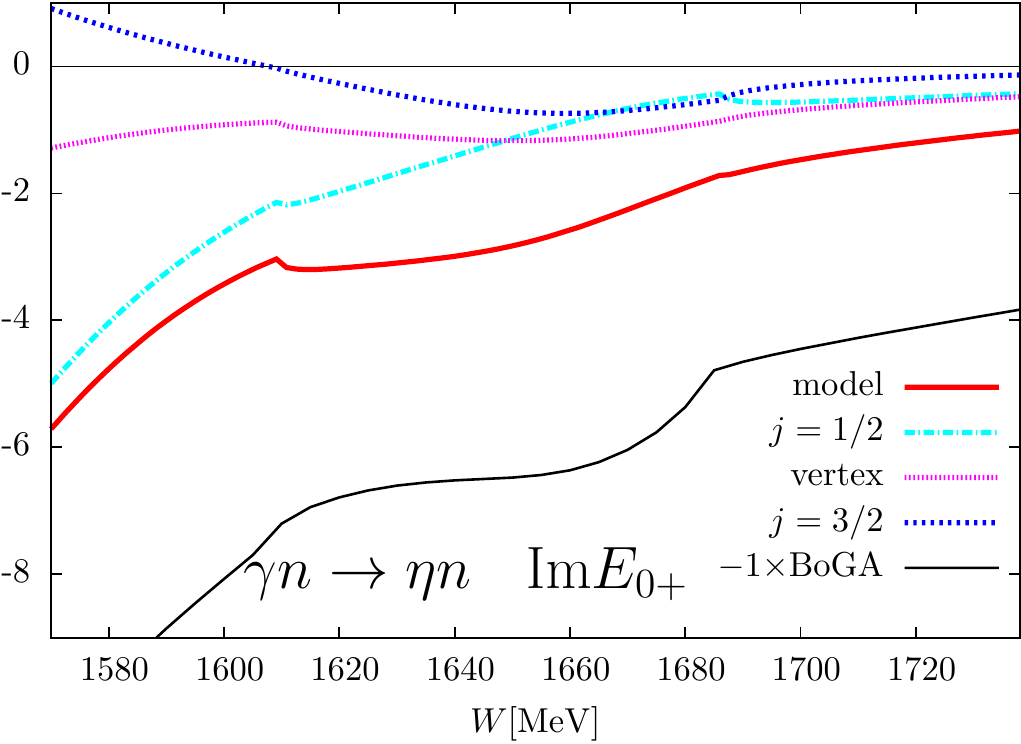}\kern2mm
\includegraphics[width=63mm]{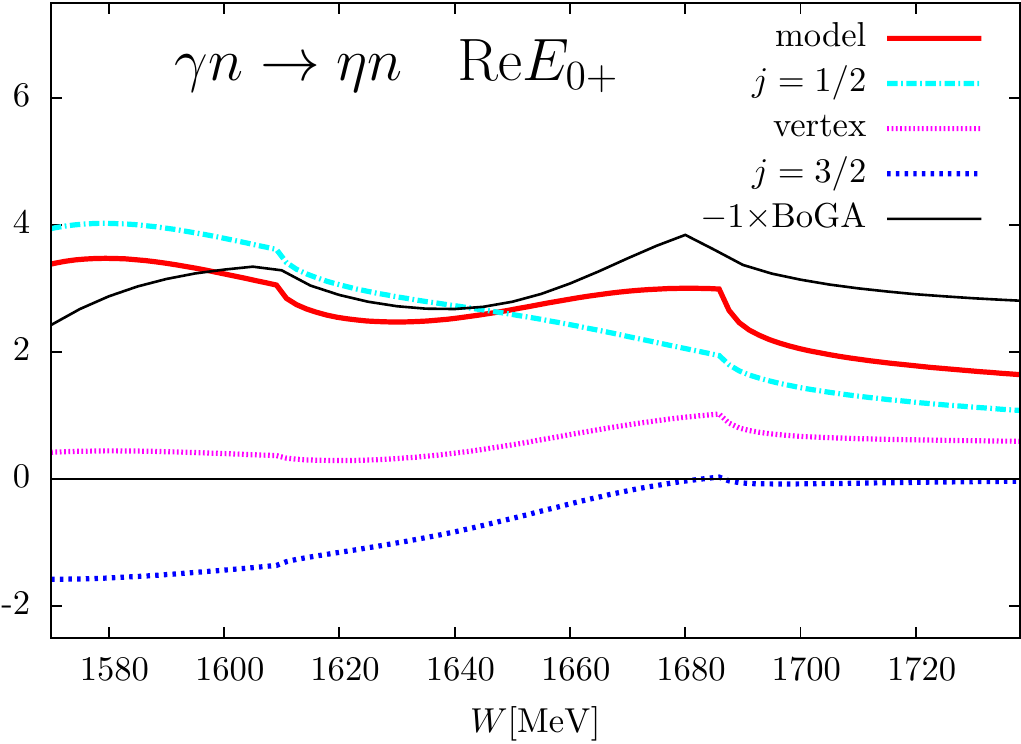}}
\caption{The dominant contributions to the imaginary and real part 
of the $E_{0+}$ amplitude (in units of mfm) for the proton 
(upper two panels) and for the neutron (lower two panels).
Apart of the separate contributions from the $s\to p_{3/2}$ and  
$s\to p_{1/2}$ transitions the vertex correction is also displayed.
The Bonn-Gatchina results are taken from the 2014-2 dataset and  
multiplied by $-1$.}
\label{fig:E0}
\end{figure}

\begin{figure}[h!]
\hbox to\hsize{
\includegraphics[height=49mm]{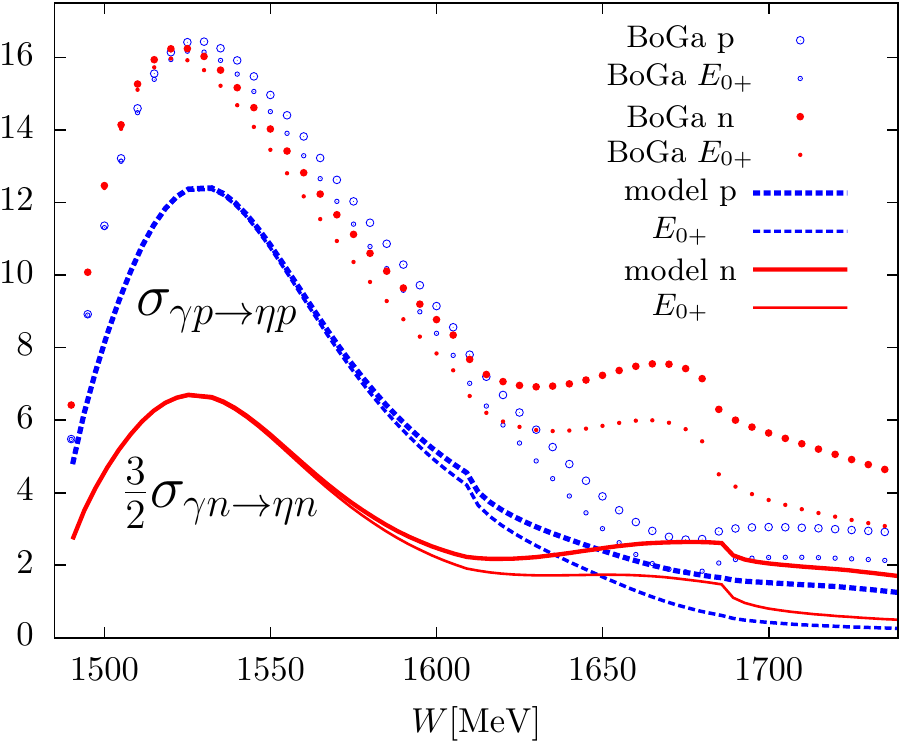}
\includegraphics[height=49mm]{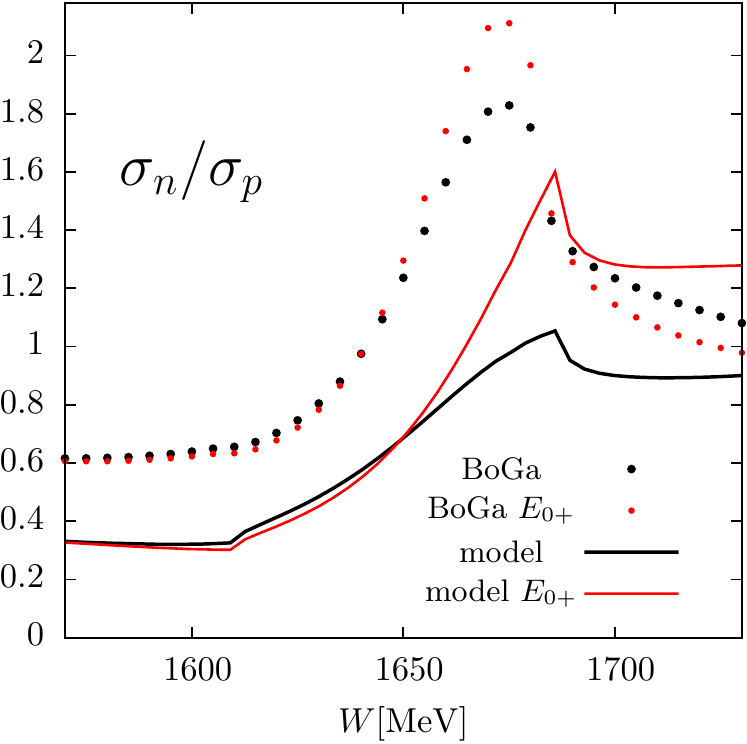}}
\caption{The total cross-sections for $\gamma p\to \eta p$ and 
$\gamma n\to \eta n$ (multiplied by the conventional factor of 
${3\over2}$) (right panel), the ratio of the neutron and the
proton cross-section (left panel).
Thinner circles and lines: contribution of the $S_{11}$ partial wave.
The BoGa curves have been reconstructed from the Bonn-Gatchina 2014-2 
data set~\cite{BoGa-data}.}
\label{fig:sigma}
\end{figure}

The $E_{0+}$ amplitudes are shown in Fig.~\ref{fig:E0} for the 
proton and the neutron in the region of the $K\Sigma$ threshold.
Our results do show a (bump-like) structure in the $\gamma n$ channel, 
which is absent in the $\gamma p$  channel, though its strength 
in the imaginary part is lower compared to the Bonn-Gatchina 2014-2 
analysis (which fits well the experimental cross-section).
A moderate rise of the neutron real amplitude below the $K\Sigma$
threshold is clearly a consequence of the contribution
from the  $j=3/2$ orbit, while the cusp-like drop in the 
amplitudes is due to the  $K\Sigma$ threshold.
This behaviour of the amplitudes is reflected in the 
cross-section as a peak (bump) present only in the neutron channel
(see Fig.~\ref{fig:sigma}).
Though the strength in our model is lower compared to the 
Bonn-Gatchina analysis, the qualitative agreement does offer a 
possible and straightforward explanation of this structure in 
terms of the quark model: a combination of a peculiar property of 
the (relativistic) wave functions of the $S_{11}$ resonances and 
the presence of the $K\Sigma$ threshold.
Let us stress that the proposed explanation of the considered peak
would not be possible in a framework of the nonrelativistic quark 
model in which the radial behaviour of the quark wave function 
depends only on the orbital momentum quantum number.

\section{Relation to the Single Quark Transition Model}

Our model can be envisioned as a version of the 
Single Quark Transition Model (SQTM) in which the photon 
interacts with a single quark in the three-quark core
and the other two quarks act as spectators. 
The general form of the SQTM operator is a product of
the boost operator and current operators~\cite{Cottingham1979}:
\begin{equation}
Bj^\lambda =\sum_{lSL}M^\lambda_{lSL}\mathcal{T}(l,S,L,\lambda)
        = \sum_{ll_zS}R^\lambda_{lSl_z}T(l,l_z,S,S_z=\lambda-l_z)\,,
\label{BjC}
\end{equation}
where
$$
M^\lambda_{lSL} = \CG{l}{l_z}{S}{\lambda-l_z}{L}{\lambda}R^\lambda_{lSl_z}\,,
\qquad
\langle l||T||0\rangle=1
\quad\hbox{and}\quad
\langle\half||T(S)||\half\rangle=\sqrt{2S+1}\,.
$$
In our approach the quark states are labeled by the total
angular momentum $j, j_z$ rather than the orbital angular
momentum and spin.
In this case it is more convenient to expand (\ref{BjC}) as
$$
   Bj^\lambda =\sum_{jlL}\mathcal{M}^\lambda_{ljL}\Sigma^{[j\h]}_{L\lambda}\,,
\qquad
    \langle lj||\Sigma^{[j\h]}_{L}||0\half\rangle = \delta_{j,l\pm\h}\,.
$$
Recoupling the angular momenta we find
\begin{eqnarray}
   \mathcal{M}^\lambda_{ljL}
&=&
                 \sum_{S=0,1}(-1)^{j+L-S-\h}\sqrt{2(2L+1)(2S+1)}\,
                 W(ljS\half;\half L) M^\lambda_{lSL}
\nonumber\\
      &=& 
        (-1)^{j+L-\h}M^\lambda_{l0l} 
      + (-1)^{j+L+\h}\sqrt{6(2L+1)}\,
                 W(lj1\half;\half L) M^\lambda_{l1L}\,,
\nonumber
\end{eqnarray}
where $W$ are the Racah coefficients.

In the case of $S11$ resonances $l=1$, and only the $E1$ multipole is
involved ($L=1$, $\lambda=1$).
In this case the coefficients (\ref{bg:M12}) and  (\ref{bg:M32}) read
\begin{eqnarray*}
  \mathcal{M}_\h =  \mathcal{M}^1_{1\h 1} 
&=& 
  -M^1_{101} + \sqrt{2}M^1_{111} 
= -e^{11}_1 + \sqrt2\,m^{11}_1\,,
\\
  \mathcal{M}_\th =  \mathcal{M}^1_{1\th 1} 
&=& 
   M^1_{101} + {1\over\sqrt{2}}\;M^1_{111} 
=  e^{11}_1 + {1\over\sqrt2}\;m^{11}_1\,,
\end{eqnarray*}
where $e^{11}_1$ and $m^{11}_1$ are the "quark electric" and 
"quark magnetic" multipole moments.
Table~1. in \cite{Cottingham1979} gives for the corresponding $E1$
baryon multipole moment of the proton and the neutron which in turn
can be related to (\ref{bg:M12}) and  (\ref{bg:M32}):
\begin{eqnarray*}
   {}_pE1 &=& 
  \sqrt{1\over3}\; e^{11}_1  - \sqrt{2\over3}\; m^{11}_1 
 = -{1\over\sqrt3}\;\mathcal{M}_\h\,,
\\
   {}_nE1 &=& 
 -\sqrt{1\over3}\; e^{11}_1  + \sqrt{2\over27}\; m^{11}_1 
 = {1\over9\sqrt3}\left[5\mathcal{M}_\h - 4\mathcal{M}_\th\right]\,, 
\end{eqnarray*}
in agreement with our conclusion that the $j=\thalf$ orbit 
contributes only in the $\gamma n\to \eta n$ channel,
which explains the different behaviour of the $\eta p$ and
$\eta n$ channels in $\eta$ photoproduction.

\end{document}